\documentclass{article}

\usepackage{microtype}
\usepackage{graphicx}
\usepackage{subfigure}
\usepackage{booktabs} %

\usepackage{url}
\usepackage{breakurl}
\usepackage[hidelinks,breaklinks]{hyperref} %

\usepackage[accepted]{icml2023}

\usepackage{amsmath}
\usepackage{amssymb}
\usepackage{mathtools}
\usepackage{amsthm}

\usepackage[ngerman,english]{babel}

\usepackage[capitalize,noabbrev]{cleveref}

\theoremstyle{plain}

\theoremstyle{definition}

\theoremstyle{remark}

\usepackage[obeyFinal, textsize=tiny, disable]{todonotes}

\usepackage[final]{changes} %
\usepackage{glossaries-extra} %
\glsdisablehyper
\usepackage[detect-all,binary-units,per-mode=symbol]{siunitx}
\usepackage{tikz}
\usetikzlibrary{arrows,arrows.meta,bending,positioning,calc,patterns,backgrounds,fit,decorations.pathreplacing,shapes.geometric}

\usepackage[frozencache,cachedir=.]{minted}

\usepackage{cleveref}
\usepackage{xcolor}
\usepackage[skins]{tcolorbox}
\usepackage{circuitikz}
\usepackage{sidecap}
\usepackage{nicefrac}
\usepackage{cprotect}
\usepackage{multirow}
\usepackage{booktabs}
\usepackage{lipsum}
\usepackage{siunitx}
\usepackage{bm}
\usepackage{import}
\usepackage{listings}
\crefname{lstlisting}{listing}{listings}
\usepackage{algorithm}

\definecolor{myolive}{RGB}{202,202,106}
\definecolor{myred}{RGB}{208,97,91}
\definecolor{myviolet}{RGB}{164,135,198}
\definecolor{myblue}{RGB}{98,145,191}

\setabbreviationstyle[acronym]{long-short}
\GlsXtrEnableEntryCounting{acronym}{1}
\newacronym{adc}{ADC}{analog-to-digital converter}
\newacronym{adex}{AdEx}{adaptive exponential integrate-and-fire}
\newacronym{afib}{AF}{atrial fibrillation}
\newacronym{ann}{ANN}{artificial neural network}
\newacronym{asic}{ASIC}{application-specific integrated circuit}
\newacronym{asicab}{\acrshort{asic} adapter \acrshort{pcb}}{\acrlong{asic} adapter \acrlong{pcb}}
\newacronym{api}{API}{application programming interface}
\newacronym{bmbf}{BMBF}{German Federal Ministry of Education and Research}
\newacronym{bptt}{BPTT}{backpropagation through time}
\newacronym{bss2}{\mbox{BSS-2}}{Brain\mbox{ScaleS-2}}
\newacronym{bss1}{\mbox{BSS-1}}{Brain\mbox{ScaleS-1}}
\newacronym{bss2os}{\gls{bss2} OS}{\gls{bss2} Operating System}
\newacronym{bss}{BSS}{BrainScaleS}
\newacronym{cdnn}{CDNN}{convolutional deep neural network}
\newacronym{cpu}{CPU}{central processing unit}
\newacronym{dfki}{DFKI}{German Research Centre for Artificial Intelligence}
\newacronym{dma}{DMA}{direct memory access}
\newacronym{dram}{DRAM}{dynamic random-access memory}
\newacronym{ecg}{ECG}{electrocardiogram}
\newacronym{fpga}{FPGA}{field-programmable gate array}
\newacronym{gbe}{GbE}{gigabit ethernet}
\newacronym{i2c}{I\textsuperscript{2}C}{Inter-Integrated Circuit}
\newacronym{ic}{IC}{integrated circuit}
\newacronym{isa}{ISA}{instruction set architecture}
\newacronym{itl}{ITL}{in-the-loop}
\newacronym{jit}{JIT}{just-in-time}
\newacronym{lvds}{LVDS}{low-voltage differential signaling}
\newacronym{lif}{LIF}{leaky-integrate and fire}
\newacronym{li}{LI}{leaky integrator}
\newacronym{mac}{MAC}{multiply–accumulate}
\newacronym{madc}{MADC}{membrane \gls{adc}}
\newacronym{mse}{MSE}{mean squared error}
\newacronym{cadc}{CADC}{columnar \gls{adc}}
\newacronym{pcb}{PCB}{printed circuit board}
\newacronym{ppu}{\acrshort{simd} \acrshort{cpu}}{\acrlong{simd} \acrlong{cpu}}
\newacronym{relu}{ReLU}{rectified linear unit}
\newacronym{rtl}{RTL}{Register Transfer Level}
\newacronym{gd}{GD}{gradient descent}
\newacronym{simd}{SIMD}{single instruction, multiple data}
\newacronym{snn}{SNN}{spiking neural network}
\newacronym{sodimm}{\mbox{SO-DIMM}}{small outline dual in-line memory module}
\newacronym{sram}{SRAM}{static random-access memory}
\newacronym{stdp}{STDP}{spike timing dependent plasticity}
\newacronym{stp}{STP}{short term plasticity}
\newacronym{rnn}{RNN}{recurrent neural network}
\newacronym{rsnn}{RSNN}{recurrent spiking neural network}
\newacronym{nasprop}{NASProp}{neuromorphic accumulative spike propagation}
\newacronym{vu}{VU}{vector unit}
\newacronym{udp}{UDP}{user datagram protocol}
\newacronym{cd}{CD}{continuous deployment}
\newacronym{ci}{CI}{continuous integration}
\newacronym{hpc}{HPC}{high-performance computing}
\newacronym{gpu}{GPU}{graphics processing unit}
\newacronym{usb}{USB}{universal serial bus}

\icmltitlerunning{Event-based Backpropagation for Analog Neuromorphic Hardware}

\begin{document}

\twocolumn[
\icmltitle{Event-based Backpropagation for Analog Neuromorphic Hardware}
\title{Event-based Backpropagation for Analog Neuromorphic Hardware}

\icmlsetsymbol{equal}{*}

\begin{icmlauthorlist}
\icmlauthor{Christian Pehle}{equal,kip}
\icmlauthor{Luca Blessing}{equal,kip}
\icmlauthor{Elias Arnold}{kip}
\icmlauthor{Eric Müller}{kip,einc}
\icmlauthor{Johannes Schemmel}{kip}
\end{icmlauthorlist}

\icmlaffiliation{kip}{Kirchhoff-Institute for Physics, Heidelberg University, Heidelberg, Germany}
\icmlaffiliation{einc}{European Institute for Neuromorphic Computing, Heidelberg University, Heidelberg, Germany}

\icmlcorrespondingauthor{Christian Pehle}{christian.pehle@kip.uni-heidelberg.de}
\icmlcorrespondingauthor{Luca Blessing}{luca.blessing@kip.uni-heidelberg.de}

\icmlkeywords{Machine Learning, ICML}

\vskip 0.3in
]

\printAffiliationsAndNotice{\icmlEqualContribution} %

\begin{abstract}
	Brain-inspired or neuromorphic computing aims to incorporate lessons from studying biological-nervous systems in the design of practical computer architectures.
While existing approaches have successfully implemented aspects of those computational principles, such as sparse spike-based computation, event-based scalable learning has remained an elusive goal in large-scale systems.
Reaching this goal is important because only then the potential energy-efficiency advantages of neuromorphic systems relative to other hardware architectures can be realized during learning.
We present our progress implementing such an event-based algorithm for learning ---the EventProp Algorithm--- using the example of the BrainScaleS-2 analog neuromorphic hardware.
Previous gradient-based approaches to learning used ``surrogate gradients'' and dense sampling of system observables or were limited by assumptions on the underlying dynamics and loss functions.
In contrast, our approach does only need spike time observations from the system while being able to incorporate other system observables, such as membrane voltage measurements, in a principled way.
This leads to a one-order-of-magnitude improvement in the information efficiency of the gradient estimate, which would directly translate to corresponding energy efficiency improvements in an optimized hardware implementation.
We present the theoretical framework for estimating gradients and results verifying the correctness of the gradient estimation, as well as experimental results on a low-dimensional classification task using the BrainScaleS-2 system.
Building on this work has the potential to enable scalable gradient estimation in large-scale neuromorphic hardware, including those using novel nano-devices, as a continuous measurement of the system state would be prohibitive and energy-inefficient in such instances.
It also suggests the feasibility of a full on-device implementation of the algorithm that would enable scalable, energy-efficient, event-based learning in large-scale analog neuromorphic hardware.

\end{abstract}

Neuromorphic or brain-inspired hardware aims to incorporate lessons and metaphors derived from biological nervous systems into the design of practical computer architectures.
Many complementary approaches exist, ranging from digital implementations to architectures involving novel nano-devices.
They have in common that they conceptualize computation as occurring in a network of processing elements ---``neurons'' exchanging messages ``spikes''--- with varying amounts of flexibility in the realizable network topology and configurability of the processing elements.
The question then becomes how to endow such a network with \emph{function}.
While there are, in principle, many approaches, some taking inspiration from local learning rules postulated in computational neuroscience a by now well-established approach is to use machine-learning techniques and, in particular, gradient-based optimization methods. 

\Acrlong{itl} gradient-based training on digital neuromorphic hardware was demonstrated on the TrueNorth system~\cite{merolla2014million, esser2016convolutional_nourl}.
To side-step the discontinuous state transition in the neuron model, a pseudo-derivative was introduced, which then resulted in a differentiable computation graph. The limited hardware weight precision was addressed by tracking a floating point precision weight in software, which was updated by gradient information during a backward pass computed on a conventional computer. 

While in digital neuromorphic hardware neuron models are numerically implemented, with typically fixed discretized time-step, analog neuromorphic hardware implements \emph{physical} neuron models, which inherently operate continuously in time.
Gradient-based hardware \acrlong{itl} %
learning has also been demonstrated to be a suitable learning scheme for analog neuromorphic hardware~\cite{arnold2022spikinghardware,cramer2022surrogate,goeltz2021fast, schmitt2017neuromorphic_nourl}.

However, these algorithms either make use of dense observations of neuron dynamics~\cite{cramer2022surrogate}, rely on a rate limit of the observed spike activity~\cite{schmitt2017neuromorphic_nourl}, or are limited by assumptions on time constants, spiking behavior, and network topology~\cite{goeltz2021fast}.

Here, we introduce a gradient-estimation algorithm, based on the EventProp algorithm~\cite{wunderlich2021event}, that  only needs spike time observations of the Neuromorphic Hardware and is suitable for arbitrary network topologies, dynamics and loss functions, eliminating all those limitations.

More specifically we make the following contributions:
\begin{enumerate}
\item We show that spike observations are sufficient for estimating gradients in analog neuromorphic hardware emulating spiking neurons. By training a feed-forward network of 120 \gls{lif} neurons on a low dimensional classification task, we demonstrate for the first time that the EventProp algorithm can be used as an \gls{itl} training algorithm for analog neuromorphic hardware.
\item By comparing with a surrogate gradient implementation both in simulation and on hardware using \gls{itl} training, we show that both algorithms solve the task with equivalent performance. The proposed algorithm is more information efficient, resulting in a ten-fold reduction in required data.
\item We demonstrate that the gradient estimation is robust to the variability of the underlying neuromorphic substrate. More specifically, we show for a test case, that the mean gradient estimate obtained using spike time hardware measurements by our algorithm agrees with the analytical solution.
\end{enumerate}
Taken together we believe this indicates that the proposed approach holds promise as an efficient and scalable solution to gradient estimation in analog neuromorphic hardware.

\begin{figure}[tb]
	\includegraphics[width=\linewidth]{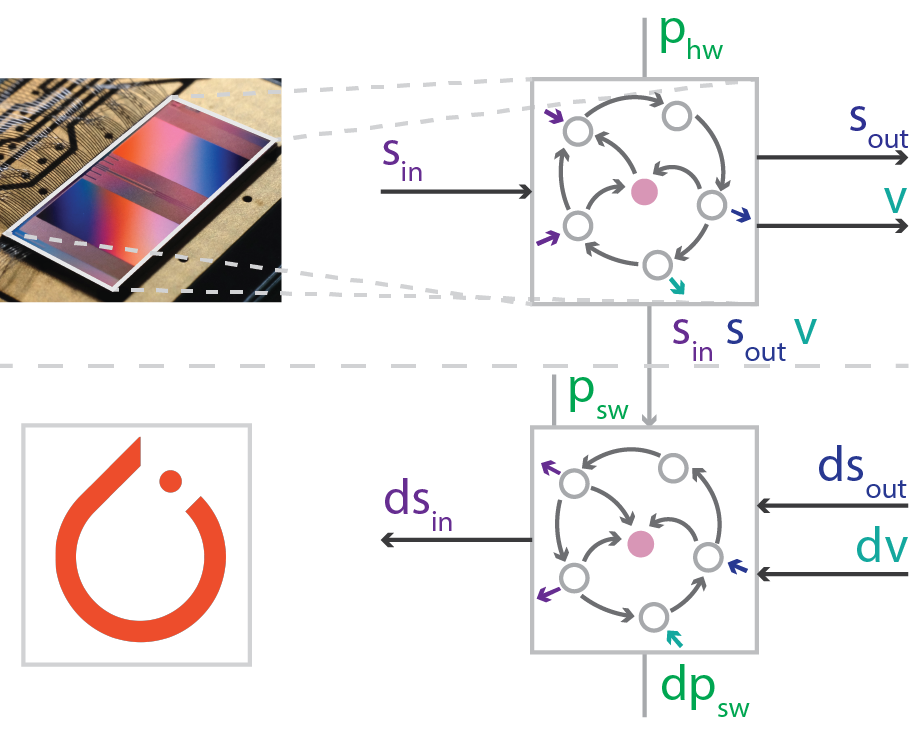}
	\caption{%
		Illustration of the \gls{itl} gradient optimisation approach for analog neuromorphic hardware. During the forward pass a spiking neural network is physically emulated on the \gls{bss2} neuromorphic chip. Hardware parameters $p_\mathrm{hw}$ setting up the digital synaptic weights of the network are  programmed and then spikes $s_\mathrm{in}$ are sent into the chip. The resulting spikes produced by the network $s_\mathrm{out}$, and optionally a subset of the membrane voltage traces $v$, are recorded. During the backward pass the recorded hardware membrane voltages $v$, input spikes $s_\mathrm{in}$ and output spikes $s_\mathrm{out}$ are used to estimate parameter gradients for the hardware system, based on a software model implemented in PyTorch.
		Top left: A \gls{bss2} neuromorphic chip (\SI{4}{\mm} \texttimes{} \SI{8}{\mm}) bonded to a carrier board, see \citet{pehle2022brainscales2_nopreprint_nourl} for detailed information.
	}%
	\label{fig:bss-2}
\end{figure}

\section{Neuromorphic Hardware}
A wide variety of neuromorphic or brain-inspired hardware architectures have been proposed, representing many complementary approaches~\cite{thakur2018mimicthebrain_nourl}. 
While digital systems rely on simulation using numerical calculations, physical models
 use analog or physical properties of a substrate for some aspect of the implemented ``neuro-inspired'' computation to gain an advantage in terms of power efficiency, speed, or density relative to digital computer architectures.
\Gls{bss2} is a research platform (\Cref{fig:bss-2}) based on a mixed-signal neuromorphic system using  conventional CMOS technology.
Its analog network core emulates 512 spiking \gls{adex} neuron circuits in continuous time, typically accelerated by $10^3$ compared to biological time scales.
Each circuit is configurable to exhibit \gls{lif} or \gls{li} dynamics and connects to a column of 256 \SI{6}{\bit} synapses. The connectivity is restricted to a definite sign per synapse row (the synapses are organized in two $256 \times 256$ arrays).
To implement signed synapses (in violation of Dale's law) two synapse rows are required.
Larger synaptic input counts or multi-compartmental neuron morphologies are realized by connecting circuits into ``logical'' neurons.
For further details, we refer to~\cite{pehle2022brainscales2_nopreprint_nourl,billaudelle2022accurate}.
On-chip digital components handle spike event communication and provide memory-mapped access to, e.g., neuron-individual parameters, synaptic weights, and other settings.
Besides allowing the implementation of flexible local learning rules or general control tasks, two embedded \gls{simd} processors can access on-chip observables such as membrane voltages via \glspl{adc} as well as off-chip memory.
A \gls{fpga} orchestrates experiments and turns the neuromorphic hardware into a network-attached \gls{snn} accelerator.
\Gls{dram} on the \gls{fpga} \gls{pcb} serves as a real-time capable playback buffer, e.g., for input stimuli or timed configuration commands, and stores streamed-out chip data.

\section{Advantages of Event-based training}
In order to implement the surrogate gradient \gls{itl} learning scheme~\cite{cramer2022surrogate}, the membrane voltage of the neuron circuits is digitized and stored with a temporal resolution of approximately \SI{2}{\micro\second}.
This imposes a significant burden both in terms of energy consumption and memory use, making a more information-efficient gradient estimation algorithm desirable.
Moreover, to compute the surrogate gradient the membrane recording is interpolated in software and a dense membrane trace is constructed for all neurons.
This leads to a computational overhead, which affects training speed.
In our experiments, we observe a speed-up in the EventProp-based training, which would further increase if the loss function was only dependent on spike times.

The external memory bandwidth and capacity can also be a limiting factor.
In the particular hardware system under consideration, up to 512 neurons can be used with each membrane sample providing an \SI{8}{\bit} value per neuron and an spike event being represented by a \SI{8}{\bit} label and a \SI{16}{\bit} timestamp.
Digitizing the membrane voltage state with a frequency of \SI{500}{\kilo\hertz}, results in data transfer rates of \SI{2}{\giga\bit\per\second}.
The external trace memory has a capacity of \SI{512}{\mebi\byte}, therefore the surrogate gradient approach accommodates at most \SI{4}{\second} of experiment runtime.
Biologically plausible firing rates of neurons are around \SI{10}{\hertz}, resulting in an average expected spike rate of \SI{10}{\kilo\hertz} on \gls{bss2}.
This translates to approximately \SI{0.12}{\giga\bit\per\second} required bandwidth and therefore one order of magnitude improvement in memory efficiency of the proposed algorithm relative to the surrogate gradient approach~\cite{cramer2022surrogate} on \gls{bss2}.
For smaller networks, or lower hardware utilization, the encoding overhead of the membrane samples increases.
Similar estimates would apply to other analog or digital neuromorphic hardware, but the details depend on the memory representation of spike and voltage data.

We can estimate the information efficiency of the proposed method compared to surrogate-gradient-based training in terms of the number of observed spike events $n_e$, the number of bits used to represent a spike event $b_e$, the number of voltage samples that are recorded $n_v$ to compute the surrogate gradients and the number of bits per voltage sample $b_v$. The surrogate gradient method needs both voltage trace data and spike observations, so it requires $n_v b_v + n_e b_e$ bits of information. In contrast the gradient estimation algorithm used in this study relies on $n_e b_e$ bits of information. Therefore the overall gain is given by
\begin{equation}
1 + \frac{n_v b_v}{n_e b_e}.
\label{eq:information_efficiency}
\end{equation}
Since the number of required voltage samples is fundamentally determined by the time constants of the membrane voltage dynamics, whereas the number of observed spike events is typically orders of magnitudes lower, this leads to the conclusion that the method is generically
more information efficient.
This estimate applies to the particular kind of neuron model under consideration. Similar estimates apply to other processing elements coupled by events, as long as they are solely coupled by messages passed between them. 

\begin{figure}

	\includegraphics[width=\linewidth]{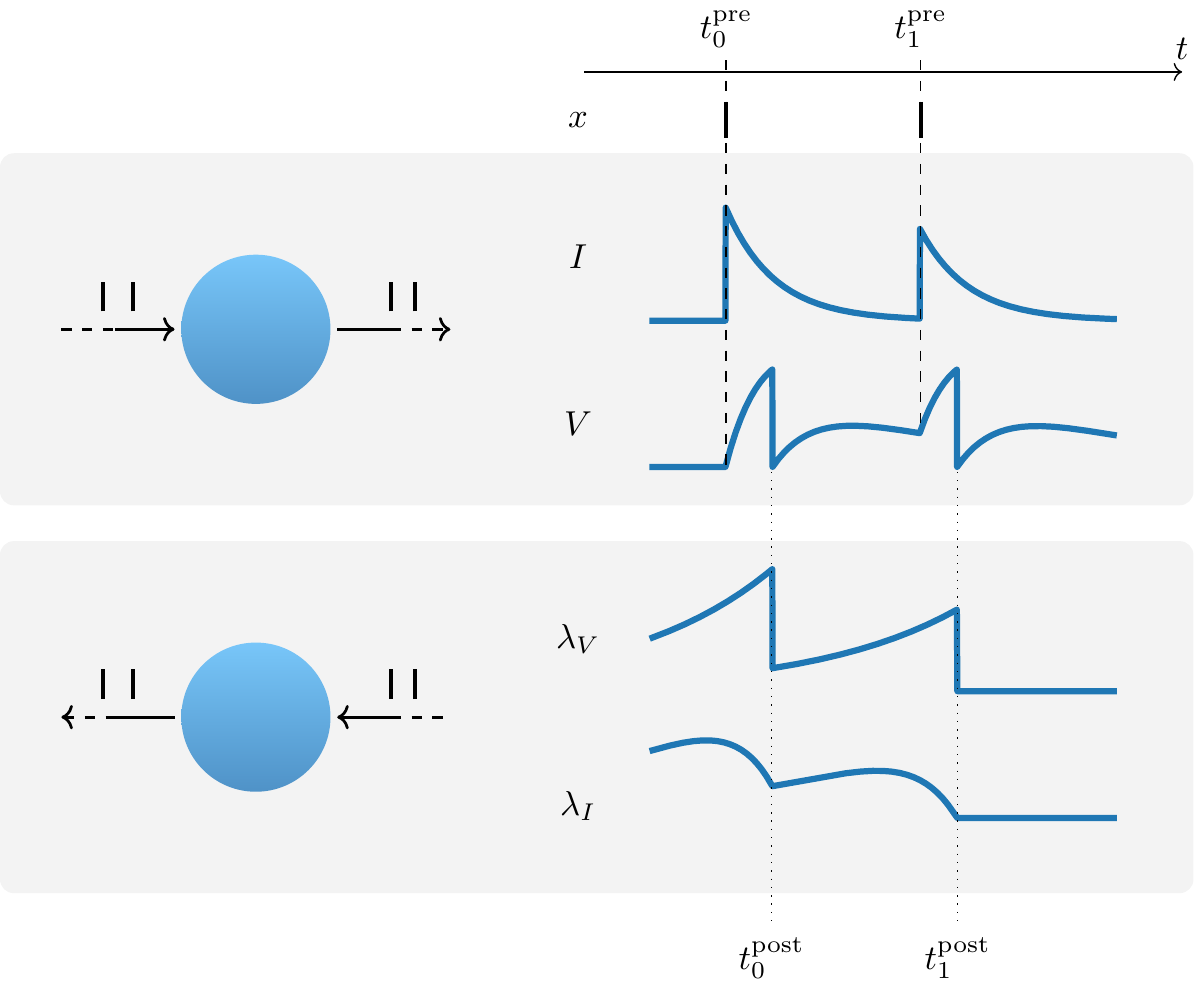}
	\caption{%
			Forward and adjoint (backward) dynamics of a LIF neuron receiving two input spikes at $t_{1}^{pre}$ and $t_{2}^{pre}$. The membrane potential experiences two jumps, giving the post-synaptic spike times $t_{1}^{post}$ and $t_{2}^{post}$. In the computation of the adjoint dynamics, which happens backwards in time, the adjoint variable $\lambda_{V}$ experiences jumps at the two post-synaptic spike times. The gradient contributions are then computed by sampling the adjoint state at the respective pre-synaptic spike times.
	}%
	\label{fig:eventprop_illustration}
\end{figure}

\section{Hardware Training Results on the Yin-Yang Dataset}

\begin{figure*}[tb]
	\centering%
	\includegraphics{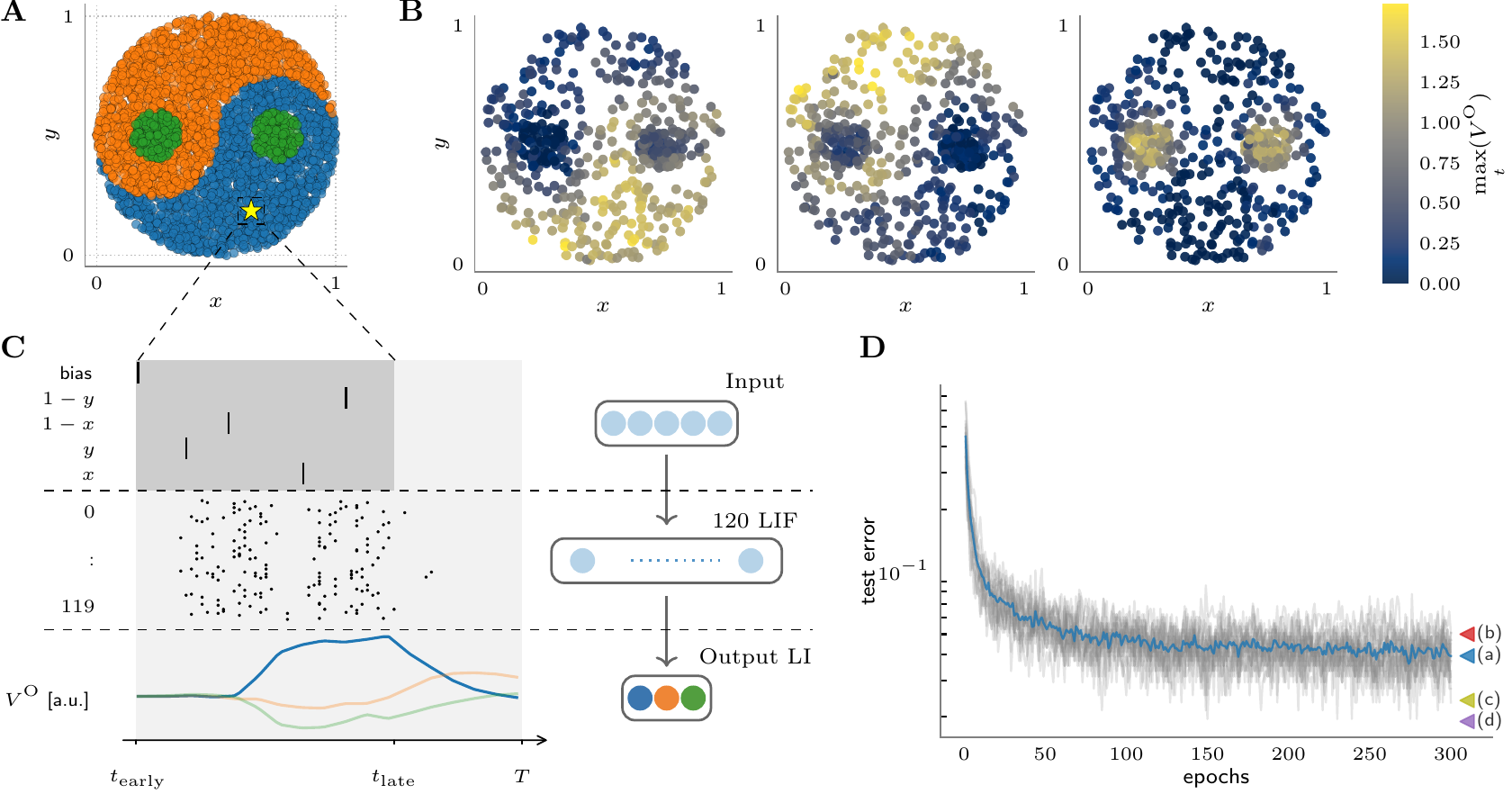}%
	\caption{%
	\textbf{A} Example points of Yin-Yang dataset~\cite{kriener2021yin}.
	The dataset is represented by two-dimensional points classified into one of the three classes yin (blue), yang (orange) or dot (green).
	\textbf{B} The maximum membrane values of the three output \gls{li}-neurons are used for classification and are shown for the test set after training.
	\textbf{C} Samples from the dataset are symmetrized and encoded into a time window $[t_{\text{early}}, t_{\text{late}}]$.
	Additionally, a bias spike is added.
	For classification, those input spikes are projected onto a hidden layer of 120 \gls{lif} neurons.
	The hidden spikes are received by an output layer of 3 \gls{li} neurons.
	The activity and observables of the network are shown for the yellow sample in A.
	Spikes and membrane voltages use different timestamp clocks; we only approximately aligned the two time domains.
	\textbf{D} The test error for hardware-\gls{itl} training with EventProp on the Yin-Yang dataset is depicted and compared to other results as given in \cref{tab:result_comparison}.
	}%
	\label{fig:training_results_yinyang}
\end{figure*}

\begin{table*}[tb]
	\centering
	\begin{tabular}{l l l l l r}
		\toprule
		\textbf{Type} & \textbf{Grad. Estimator} & & & \textbf{Loss} & \textbf{Acc. [\%]} \\
		\midrule
		ANN & Backprop. \cite{kriener2021yin} & & & CE & 97.6 \textpm 1.5 \\
		 & Backprop. \cite{kriener2021yin} & & & \multirow{2}{*}{CE} & \multirow{2}{*}{85.5 \textpm 5.8} \\
		 & \multicolumn{3}{l}{(frozen lower weights)} & & \\
		\midrule
		SNN & analytical \cite{goeltz2021fast} & sim. & & TTFS & 95.9 \textpm 0.7 \\
		 & analytical \cite{goeltz2021fast} & hw & \begingroup \color{myred} $\blacktriangleleft$ \endgroup (b) & TTFS & 95.0 \textpm 0.9 \\
		 & EventProp \cite{wunderlich2021event} & sim. & \begingroup \color{myviolet} $\blacktriangleleft$ \endgroup (d) & TTFS & 98.1 \textpm 0.2 \\
		\midrule
		SNN & Surr. Gradient & sim. & & Max & 97.6 \textpm 0.4 \\
		 & Surr. Gradient & hw & & Max & 94.6 \textpm 0.7 \\
		 & EventProp & sim. & \begingroup \color{myolive} $\blacktriangleleft$ \endgroup(c) & Max & 97.9 \textpm 0.6 \\
		 & EventProp & hw & \begingroup \color{myblue} $\blacktriangleleft$ \endgroup (a) & Max & 96.1 \textpm 0.8 \\
		\bottomrule
	\end{tabular}
	\caption{%
		Test accuracies on the Yin-Yang task for \gls{ann}s and \gls{snn}s with different gradient estimators and loss definitions.
		For the \gls{snn}s numerical integration and hardware emulation are compared.
		The results marked by `CE' use a cross-entropy loss, the `TTFS' results are using a loss based on time-to-first-spike decoding and the `Max' results are based on `maximum membrane value over time'.
		\Citet{kriener2021yin} use an \gls{ann} with 30 hidden neurons. \Citet{wunderlich2021event} use a \gls{snn} with 200 hidden neurons, the results by \citet{goeltz2021fast} and our own results use 120 hidden neurons.
	}%
	\label{tab:result_comparison}
\end{table*}

We choose the Yin-Yang dataset~\cite{kriener2021yin}, a two-dimensional classification task with three classes, to demonstrate the learning algorithm on \mbox{BrainScaleS-2}. %
A point $(a,b)$ in the dataset lies in the interior of the box $[0,1] \times [0,1]$ and belongs to one of three classes depending on which part of the ``Yin-Yang'' it belongs to.
In Table~\ref{tab:result_comparison} and Fig.~\ref{fig:training_results_yinyang} we report our results.
Hyperparameters and training details are given in Section~\ref{subsec:training_details} and Table~\ref{tab:training_parameters}.

We compare our surrogate-gradient and EventProp implementation both in simulation and on hardware in-the-loop training.
We find that in simulation both the surrogate gradient and EventProp implementation achieve comparable performance. Using the hardware-in-the loop approach to estimate gradients we achieve on average $1.5\%$ higher accuracy with the EventProp gradient estimation method and the highest observed hardware performance on this task so far, compared to previously reported results using the Fast-and-Deep gradient estimation algorithm~\cite{goeltz2021fast}.

We use Eq.~\eqref{eq:information_efficiency} to estimate the gain in information efficiency for this particular experiment on the Yin-Yang dataset. Sampling the membrane voltages of 120 hidden neurons with \SI{500}{\kilo\hertz} over \SI{38}{\micro\second} results in a total of $n_v=2280$ voltage samples per input datapoint. In the last epoch of the training with EventProp, we measure an average of $n_e = 146 \pm 13$ spikes in the hidden layer per input datapoint of the training set. This leads to a factor of $6.2 \pm 0.5$ improvement in memory efficiency using EventProp compared to training with surrogate gradients.

\section{Hardware Gradient Estimation}

\begin{figure}[tb]
	\centering
	\definecolor{dataflow_blue}{HTML}{d5ecfd}  %
\definecolor{dataflow_grey}{HTML}{d5d5d5}  %
\definecolor{upper_blue}{HTML}{cbeaff} %
\definecolor{lower_blue}{HTML}{b6d3e8} %
\definecolor{full_upper_blue}{HTML}{78c6f9} %
\definecolor{full_lower_blue}{HTML}{4f92c7} %

\begin{tikzpicture}

\node[circle, minimum size=1.5cm, top color=upper_blue, bottom color=lower_blue] (left_circle) {};

\node[right=1cm of left_circle, circle, minimum size=1.5cm, top color=full_upper_blue, bottom color=full_lower_blue] (mid_circle) {};

\node[right=1cm of mid_circle, minimum height=0.8cm, minimum width=1.7cm, rounded corners, top color=upper_blue, bottom color=lower_blue] (loss) {$L = t_{0}^{\mathrm{post}}$};

\draw[->, thick] (left_circle.east) -- (mid_circle.west) node[below, midway] (weight) {$w$};

\draw[->, thick] (mid_circle.east) -- (loss.west);

\end{tikzpicture}
	\begin{center}
		\resizebox{\linewidth}{!}{\input{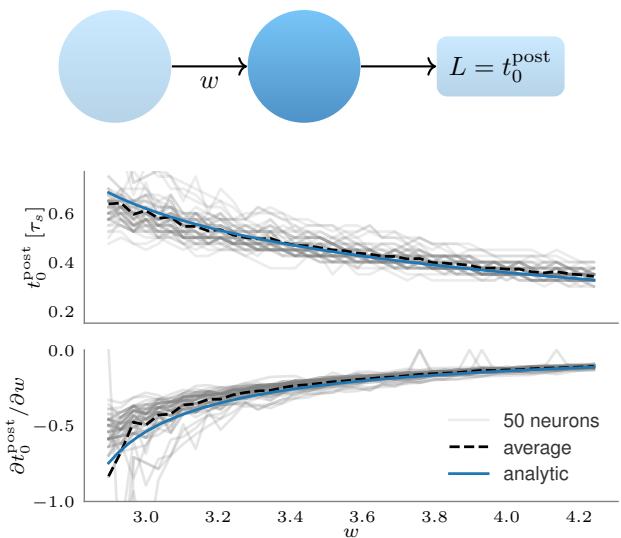}}
	\end{center}
	\caption{%
		Experiment setup (top), spike time (middle) acquired from runs on BrainScaleS-2 and estimated gradient (bottom) for a loss function $L = t^\mathrm{post}$ as a function of the weight $w$. %
		Results for $n = 50$ hardware neuron circuits are shown in light grey, the average in dashed black. The analytical result (blue line, see Fig.~\ref{fig:sw_gradients_decreasing_dt}) agrees well with the hardware average.
		}%
		\label{fig:correct_gradients_plot}
\end{figure}

Since previous work~\cite{goeltz2021fast} had demonstrated that gradient-estimation using an analytical formula for the spike time of LIF neurons could successfully be applied to hardware in-the-loop training, we ask whether the gradient estimate computed using our method would match the analytic estimate.
We find that for a simple experiment setup with one LIF neuron receiving one input spike with weight $w$ at time $t = 0$, the mean of the estimated gradient of the loss function $L = t^\mathrm{post}$, where $t^\mathrm{post}$ is the spike time of the LIF neuron, agrees well with the analytical prediction (see Fig.~\ref{fig:correct_gradients_plot}).

\section{Discussion}\label{sec:discussion}

We demonstrated a gradient estimation algorithm for analog neuromorphic hardware requiring only spike observations and making no assumptions on the network topology or loss function.
As such, this has the potential to enable scalable gradient estimation in large-scale neuromorphic hardware since continuous measurement of the system state would be prohibitively expensive in this case.
In particular, a similar approach to the one pursued here would also apply to neuromorphic systems for which continuous sampling of the system state would be infeasible even in principle, such as ones based on novel nano-devices or photonic neuromorphic systems. 
While the original implementation of EventProp~\cite{wunderlich2021event} relied on a custom event-based solver, we use here a PyTorch implementation and time-discretized forward- and adjoint dynamics.
Such a time-discretized implementation would also be suitable for digital neuromorphic hardware.
Future work will take advantage of an event-based solver to demonstrate tasks that require precise spike timing or computation on sparse data.

Although the results reported here are encouraging, further work is needed in several areas:
\begin{itemize}
	\item We would like to demonstrate the algorithm on further tasks, particularly ones that are not feasible using surrogate-gradient-based in-the-loop training due to hardware trace-memory limitations.
	\item We would like to demonstrate the scalability of the algorithm by applying it on larger-scale networks, and on longer time scales.
	\item We would like to demonstrate the algorithm on other neuron configurations supported by the chip. 
\end{itemize}

Going beyond the immediate applications of this work, we would like to use hardware observables to learn the dynamics, instead of assuming a particular model.
Demonstrating the scalability of the algorithm beyond the  modest size of the network considered here requires both a multi-chip experiment setup, which is currently being built, and potentially network models that account for delays.
We have evaluated the EventProp algorithm itself on larger-scale convolutional feed-forward architectures with $\sim 10^5$ neurons and $\sim 10^6$ parameters and found no performance disadvantage over surrogate gradients.
However, we were fundamentally limited by the small number of integration timesteps and memory constraints.

In digital neuromorphic hardware, it is relatively straightforward to simulate the implemented neuron dynamics with commodity hardware.
Thus, this work can be viewed mainly as a demonstration that the temporal discretization and quantization of observables are not an obstacle to the implementation of the EventProp gradient estimation algorithm.

Our results also suggests that a fully on-device implementation of the algorithm would be possible, where the required spike time information and voltage slopes at spike times could be stored locally at each neuron circuit. Both the forward and adjoint dynamics can be realised within the analog computation paradigm and in the current form with the very similar analog circuits. Following this direction would enable scalable, energy-efficient, event-based learning in large-scale analog neuromorphic hardware.

\section{Methods}\label{sec:methods}

\subsection{Adjoint Sensitivity Analysis with Jumps}
The BrainScaleS-2 system can emulate the \gls{adex} neuron model~\cite{gerstner2009adex} faithfully~\cite{billaudelle2022accurate}. To illustrate how to determine paramter gradients we consider here only an \emph{adaptive} leaky-integrate and fire neuron. Its equations specify a hybrid-ordinary differential equation and we use adjoint-sensitivity analysis with jumps to derive exact gradients for this model, following the approach of \citet{wunderlich2021event}. We are considering a loss function that can depend on the spike times and voltage 
traces 
\begin{align}
L = l_p(t^p) + \int_0^T l(V) \mathrm{dt}
\end{align}
more general loss functions can also be considered and indeed the max-time-loss does not follow this form. For a derivation of how the max-over-time loss can be incorporated see \citet{wunderlich2021event}. The dynamical equations
of the model are given by
\begin{align}
C \dot{V} &= -g_{L} (V - E_L) - w + R I \\
\tau_w \dot{w} &= a (V - E_L) - w \\
\tau_s \dot{I} &= -I. \label{eq:current_dynamics}
\end{align}
Here $C$ is the membrane capacitance, $g_L$ the leak conductance, $E_L$ the leak potential, $w$ the adaptation variable, $R$ the input resistance and $\tau_w, \tau_s$ are the adaptation and synaptic time constants respectively.
After a change of variables $v = V - E_L$, we can write these equations in matrix form
\begin{gather}
\label{eq:forward_dynamics}
    \dot{x} = \begin{bmatrix} \dot{v} \\ \dot{w} \\ \dot{I} \end{bmatrix}
    = A x = 
     \begin{bmatrix}
      -\tau^{-1}_{m} & -1/C & R/C \\
      -a \tau^{-1}_w & -\tau^{-1}_w & 0 \\
      0         & 0         & -\tau^{-1}_s
      \end{bmatrix}
	  \begin{bmatrix} v \\ w \\ I \end{bmatrix}.
\end{gather}
When $v$ reaches a threshold $\vartheta$, then the neuron is subject to the following transition in its state:
\begin{gather}
    \begin{bmatrix} v^+ \\ w^+ \\ I^+ \end{bmatrix}
    =
     \begin{bmatrix}
      0 & 0 & 0 \\
      0 & 1 & 0 \\
      0 & 0 & 1 \\
      \end{bmatrix}
	  \begin{bmatrix} v^- \\ w^- \\ I^- \end{bmatrix} + \begin{bmatrix} v_r \\ b \\ 0 \end{bmatrix}.
\end{gather}
If we consider a network of $n$ neurons with state vector $x$, connected by a synaptic weight matrix, then if neuron $i$ reaches its threshold the resulting transition can be written as
\begin{gather}
	x^+ = P^T_i T P_i x^- + (\mathbb{1} - P^T_i P_i) x^- + p_i, \label{eq:neuron_network_transition}
\end{gather}
with $P_i$ the projection of the $3 n$ dimensional state space to the $3$ dimensional state-space of neuron $i$,
\begin{gather}
T = \begin{bmatrix}
        0 & 0 & 0 \\
        0 & 1 & 0 \\
        0 & 0 & 1 \\
    \end{bmatrix}
\end{gather}
and $p_i$ given by the translation induced by synaptic transmission, the reset of the membrane potential and shift of the adaptation constant.

Based on this description it is easy to compute the associated adjoint sensitivity equations with jumps, as well as the corresponding parameter gradients. The adjoint equations are given by,
\begin{gather}
\lambda'
=
 \begin{bmatrix}
  -g_L/C &   -a/\tau_w &  0\\
  -1/C   & -1/\tau_w   & 0 \\
   R/C   & 0           & -1/\tau_s
 \end{bmatrix} \lambda ,
\end{gather}
and their jumps are computed according to
\begin{align}
	(\lambda^-)^T = (\lambda^+)^T \left[[f^+ - \partial_{x^-} \theta f^-] \frac{\partial_{x^-} j}{\partial_{x^-} j \dot{x}^-} + \partial_{x^-} \theta\right].
\end{align}
Here we have
\begin{align}
	f^+ &= A (P^T_i T P_i x^- + (\mathbb{1} - P^T_i P_i) x^- + p_i), \\
	f^- &= A x^-, \\
	\partial_{x^-} \theta &= P^T_i T P_i + (\mathbb{1} - P^T_i P_i), \\
	\partial_{x^-} j &= \partial_{x^-} (a^T x^- + w) = a^T,
\end{align}
where $A$ was defined in Eq.~\eqref{eq:forward_dynamics} and $a^T = (1,0,0)$. This simplifies the equation to
\begin{align}
(\lambda^-)^T = (\lambda^+)^T \left[A p_i \frac{a^T}{a^T \dot{x}^-} + P^T_i T P_i + (\mathbb{1} - P^T_i P_i)\right].
\end{align}
There exists an equally simple expression for the gradient contributions to the synaptic weights $w_{ji}$~\cite{wunderlich2021event}
\begin{align}
	\frac{dL}{dw_{ji}} = -\tau_s\sum_{\mathrm{spikes \, from \, i}} (\lambda_I)_j.
\end{align}
The formulas derived here simplify to the ones obtained in~\cite{wunderlich2021event}, if the adaptation variable $w$ is omitted.
It is also possible to use the same overall approach to derive explicit adjoint equations for the full AdEx neuron model, including the case of an absolute refractory period or synaptic transmission delays. One consideration is that since the full AdEx dynamics is non-linear, the full membrane voltage trace enters the adjoint dynamics. This introduces an additional complication in the potential applicability of the method.

\subsection{Software Framework}

Our software stack translates the high-level \gls{snn} experiment description to a data flow graph representation, places and routes neurons and synapses on the hardware substrate, and compiles stimulus inputs, recording settings, and other runtime dynamics into an experiment program representing an equivalent experiment configuration on BrainScaleS-2, see \citet{mueller2022scalable} for a detailed description.
The analog substrate on BrainScaleS-2 is subject to device variations, or fixed-pattern noise, that can be compensated for by calibration.
At the same time, the calibration routines consider user-defined model parameters to provide an equivalent parametrization of the emulation.
A complete calibration data set provides per-circuit operation point settings.
The training employed in this paper only affects the digital weight parameters
and is implemented as an incremental reconfiguration providing quick hardware-\gls{itl} updates.
The BrainScaleS-2 hardware substrate only supports fixed-sign \SI{6}{\bit} synapses.
We allocate two hardware synapses per software weight $w_\text{sw}$ to support efficient signed weight matrix updates.
Each software weight $w_\text{sw}$ is linearly scaled into a hardware-compatible range and rounded to the nearest integer value.
The batched input spikes are injected into BrainScaleS-2 and the \gls{snn} is emulated for $T = \SI{38}{\micro\second}$ per batch entry.
During emulation, spike events and neuron membrane traces can be recorded to the \gls{fpga} \gls{dram}.
The host computer reads back and post-processes the recorded data.
For experiments relying on \gls{cadc} membrane measurements, the membrane samples are expressed on an equidistant time grid by linear interpolation in order to be represented by a \texttt{torch::Tensor} with a fixed time grid and thus aligned to the PyTorch API\@.
Additionally, the values are offset and scaled into a desired range.
The spike recordings are mapped to a boolean tensor on the same time grid.

This hardware-\gls{itl} operation mode on BrainScaleS-2 is supported by \texttt{hxtorch.snn}~\cite{spilger2022hxtorchsnn}, a \texttt{PyTorch}-based~\cite{paszke2017automatic} library that automates and abstracts away hardware-specific procedures and provides data conversions from and to \texttt{PyTorch}.

\subsection{Numerical Gradient Estimate}

We discretize the forward and adjoint dynamics using the explicit Euler integration scheme to implement the EventProp Algorithm, which is derived in continuous time~\cite{wunderlich2021event}. The dynamics of the LIF neurons with exponential-shaped, current-based synapses are either computed in simulation only or can be injected from observations when training with hardware in the loop. This, together with computing the adjoint trajectories, is handled in a custom \texttt{torch.autograd.Function}. The complete dataflow, together with another function ensuring the correct backpropagation to the synaptic weights and the previous layer is displayed in \cref{fig:pytorch_modules}. The gradient estimation for a layer of LIF neurons are described in \Cref{alg:discrete_eventprop}.

For the backward computation we need the spike times and the time derivatives of the membrane $\dot{V}$, which are determined only by the synaptic currents $I$ at spike times. When training with hardware in the loop, synaptic currents are not accessible and need to be estimated in order to be able to calculate the jumps of the adjoint variables. We estimate the synaptic currents by numerically integrating \cref{eq:current_dynamics} while assuming ideal dynamics on hardware and use the boolean tensors, to which the spike recordings were mapped, to apply the transitions as in \cref{eq:neuron_network_transition}.

To compare our gradient estimate from simulation to the analytical gradient, we consider the experiment setup of a LIF neuron receiving a single spike with weight $w$ as in \cref{fig:correct_gradients_plot}. For the special case of $\tau_\text{syn} = \tau_\text{mem}$, $\tau_\text{refrac} = 0$ the explicit formula from \citet{goeltz2021fast} is used.

\begin{figure}[tb]
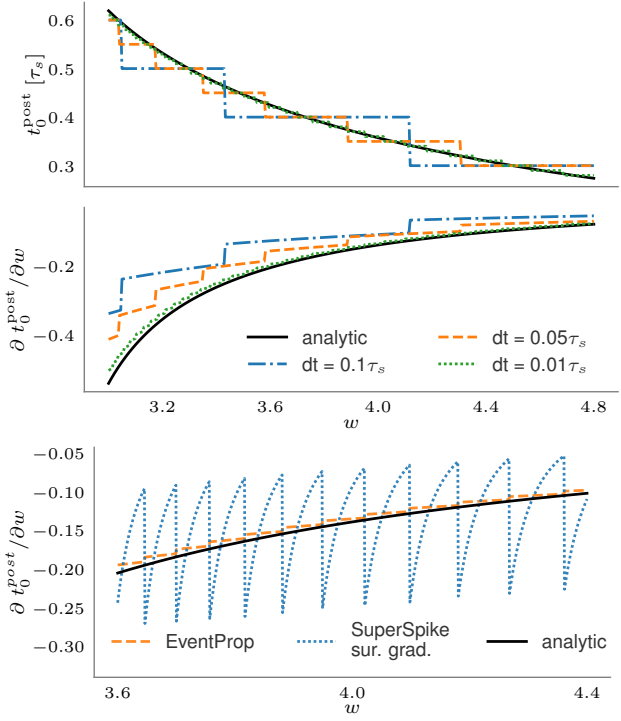

	\centering
	\begin{center}
		\resizebox{\linewidth}{!}{\input{plots/ep_sw_decreasing_dts.pgf}}
		\resizebox{\linewidth}{!}{\input{plots/ep_sg_sw.pgf}}
	\end{center}
	\caption{%
		Spike time (top) and estimated gradient (bottom) for a loss function $L = t^\mathrm{post}$ as a function of the weight $w$.
		As the numerical integration timestep $\mathrm{dt}$ decreases the numerically estimated gradient (dashed lines) converges to analytically known gradient (solid-black line), cf.\ \citet{goeltz2021fast}.
		Jumps in the gradient and spike time are due to the finite resolution $\mathrm{dt}$ of the time-grid, note that the gradient is changing although the spike time remains constant.
		}%
		\label{fig:correct_gradients_scheme}%
		\label{fig:sw_gradients_decreasing_dt}
\end{figure}

For spike time dependent losses, as for the gradient comparisons in \cref{fig:correct_gradients_plot} and \cref{fig:sw_gradients_decreasing_dt}, we need to extract spike times from dense boolean tensors.
To be able to optimize on such losses, we wrote a custom \texttt{torch.autograd.Function}, see \cref{lst:to_spike_times}.
The gradient with respect to the time of a spike is backpropagated at the index, at which the spike was located in the dense boolean input tensor.

There are two sources of errors introduced by numerical implementation: Since we only evaluate spike times on a fixed time grid, the spike time changes in jumps as the synaptic weight increases. The second error source is the numerical error introduced by the integration method itself. The chosen integration scheme is only first-order accurate. As seen in \cref{fig:correct_gradients_scheme}, the jumps in the spike times have a larger impact on the mismatch in the gradient estimate, relative to the analytical formula.

For decreasing integration time steps $dt$ the numerical gradient estimate converges to the analytically known gradient. Therefore, we consider forward euler integration sufficient for our experiments.

\begin{algorithm}[H]
	\caption{Discrete EventProp gradient estimation for a single layer of LIF neurons}\label{alg:discrete_eventprop}
	\begin{algorithmic}[0]
	\INPUT
		\STATE pre-synaptic spikes $\textbf{z}_{\mathrm{pre}}$,
		\STATE parameters $\bm{W}$, $\tau_{\mathrm{m}}$, $\tau_{\mathrm{s}}$, $dt$, $T$,
		\STATE optional: hardware data $(\textbf{V}_{\mathrm{hw}}, \textbf{z}_{\mathrm{hw}})$
	\STATE \texttt{\\}
	\COMMENT{Forward pass}
	\IF{no hardware data}
	\STATE $(\textbf{V}_{\mathrm{sw}}, \textbf{I}_{\mathrm{sw}}, \textbf{z}_{\mathrm{sw}})$ by forward euler integration of forward dynamics and jumps from $0$ to $T$, step size $dt$	
	\ELSE
	\STATE Estimate $\textbf{I}_{\mathrm{hw}}$ from $\bm{W}$ and $\textbf{z}_{\mathrm{pre}}$
	\ENDIF
	\STATE \texttt{\\}
	\COMMENT{Backward pass}
	\STATE $(\bm{\lambda}_{I}, \bm{\lambda}_{V})$ by forward euler integration of adjoint dynamics and jumps from $T$ to $0$, step size $dt$
	\STATE
	\OUTPUT $\frac{\mathrm{d} L}{\mathrm{d} \bm{W}} = -\tau_{\mathrm{s}} \bm{\lambda}_{I}^{\top} \textbf{z}_{\mathrm{pre}}$, $\frac{\mathrm{d} L}{\mathrm{d} \textbf{z}_{\mathrm{pre}}} = (\bm{\lambda}_{V} - \bm{\lambda}_{I} )\bm{W}$
	\end{algorithmic}
\end{algorithm}

\subsection{Training Details}%
\label{subsec:training_details}

The four values $(a, b, 1-a, 1-b)$ for each sample of the Yin-Yang dataset are translated into spike times using a latency encoding inside the time interval $[t_{\text{early}}, t_{\text{late}}]$. Additionally a bias spike is added at $t_{\text{bias}}$, which serves as a fifth input spike and is the same for every data point. Those times are then mapped as events onto a boolean tensor on the desired time grid.

The feed-forward network consists of a hidden layer with 120 LIF neurons and an output layer of 3 LI neurons, one for each class. The loss we use is composed of two terms $L = L_{1} + L_{2}$.
The first and main term is a max-over-time loss
\begin{align}
	L_{1} &= - \frac{1}{N_{\text{batch}}} \sum_{n=1}^{N_{\text{batch}}} \log \frac{\exp \left(\max_{t} V^{\text{O}}_{n, y_{n}} (t)\right)}{\sum_{c=1}^{C} \exp\left(\max_{t} V^{\text{O}}_{n, c} (t)\right)},
\end{align}
with time $t$, the voltages of the output layer neurons $V^{\text{O}}$, the target $y$, the batch size $N_{\text{batch}}$ and the number of classes $C$.
Additionally, to prevent the amplitudes from being to high, we add a regularization term
\begin{align}
	L_{2} &= \alpha \cdot \frac{1}{N_{\text{batch}} \, C} \sum_{n=1}^{N_{\text{batch}}} \sum_{c=1}^{C} \left(\max_{t} V_{n}^{\text{O}} (t) \right)^2,
\end{align}
where $\alpha$ is a scaling factor to adjust the influence of this amplitude regularization.

When using BrainScaleS-2 \gls{itl} training, we repeat the input per sample five times, ultimately giving us 25 input streams per data point.
The hidden weights are initialized as an $n_{\text{hidden}} \times 5$ matrix and are then also repeated 5 times along the input dimension, resulting in a $n_{\text{hidden}} \times 25$ weight matrix.
This gives us an equivalent of an increase in synaptic efficacy without changing the target model parameters and underlying calibration data set.

Comparing the gradients of the output layer weights to those of the hidden layer weights, the gradients differ by multiple orders of magnitude. To counteract this, we scale the the weight gradients computed by the EventProp algorithm with a factor $1 / \tau_{s}$, which was necessary to obtain the results on the Yin-Yang dataset.

\section*{Author Contributions}\label{sec:author_contributions}

\textbf{CP}: Conceptualization, formal analysis, writing --- original draft, writing --- reviewing \& editing;
\textbf{LB}: Validation, formal analysis, investigation, visualization, writing --- original draft, writing --- reviewing \& editing;
\textbf{EA}: Methodology, software, resources, writing --- original draft, writing --- reviewing \& editing;
\textbf{EM}: Methodology, software, resources, writing --- original draft, writing --- reviewing \& editing, supervision;
\textbf{JS}: Supervision, funding acquisition, writing --- reviewing \& editing.

\section*{Acknowledgements}\label{sec:acks}
The authors wish to thank P.\ Spilger and C.\ Mauch for software enhancements and platform operation,
S.\ Billaudelle,  J.\ G{\"o}ltz, and J.\ Weis for fruitful discussions and hardware parameterization knowledge,
and all present and former members of the Electronic Vision(s) research group contributing to the BrainScaleS-2 neuromorphic platform.

\section*{Funding}
This work has received funding from
the EC Horizon 2020 Framework Programme
under grant agreements
785907 (HBP SGA2) %
and
945539 (HBP SGA3), %
the \foreignlanguage{ngerman}{Deutsche Forschungsgemeinschaft} (DFG, German Research Foundation) under Germany’s Excellence Strategy EXC 2181/1-390900948 (the Heidelberg STRUCTURES Excellence Cluster),
the German Federal Ministry of Education and Research under grant number 16ES1127 as part of the \foreignlanguage{ngerman}{\emph{Pilotinnovationswettbewerb `Energieeffizientes KI-System'}},
the Helmholtz Association Initiative and Networking Fund [Advanced Computing Architectures (ACA)] under Project SO-092,
as well as from the Manfred Stärk Foundation,
and the \foreignlanguage{ngerman}{Lautenschläger-Forschungspreis} 2018 for Karlheinz Meier.

\bibliography{vision}
\bibliographystyle{icml2023}

\newpage
\appendix
\onecolumn
\section{Neuromorphic Architectures}
\begin{table*}[htbp]
	\centering
	\begin{tabular}{l l l}
		\toprule
		\textbf{Arch}  	                                                                        & \textbf{Hardware Model} & \textbf{Modeling Paradigm} \\
		\midrule
		SpiNNaker 1																				& soft digital				& bio.\ SNN; distributed programming \\
		\cite{furber2012overview}																& \multicolumn{2}{l}{\cite{rhodes2018spynnaker,rowley2019spinntools,galluppi2015framework,galluppi2012hierachical}} \\
		\\
		SpiNNaker 2 																			& soft digital				& bio.\ SNN; distributed programming; ANN \\
		\cite{mayr2019spinnaker}																& \multicolumn{2}{l}{\cite{rostami2022eprop,yan2019efficient}} \\
		\midrule                                                                                
		Loihi 1        						                                            		& flexible digital        & bio.\ \& ML-friendly SNN; ANN \\
		\cite{davies2018loihi}																	& \multicolumn{2}{l}{\cite{rueckauer2021nxtf,dewolf2020nengo,lin2018programming}} \\
		\\
		Loihi 2																					& flexible digital        & bio.\ \& ML-friendly SNN; ANN \\
		\cite{orchard2021efficient}																& \multicolumn{2}{l}{\cite{loihi2021lava}} \\
		\midrule                                                                                
		TrueNorth																				& hard digital            & bio.\ SNN; ANN \\
		\cite{merolla2014million}																& \multicolumn{2}{l}{\cite{amir2013cognitive}} \\
		\\
		ODIN																					& hard digital            & bio.\ SNN \\
		\cite{frenkel2018online}																& \multicolumn{2}{l}{\cite{frenkel2018online}} \\
		\\
		Tianjic																					& hard digital            & bio.\ SNN; ANN \\
		\cite{pei2019tianjic}																	& \multicolumn{2}{l}{\cite{ji2016neutrams}} \\
		\midrule                                                                                
		\acrlong{bss1}																			& physical                & bio.\ SNN \\
		\cite{schemmel2010iscas}																& \multicolumn{2}{l}{\cite{mueller2020bss1_nourl}} \\
		\\
		\acrlong{bss2}																			& physical                & bio.\ and ML-friendly SNN; ANN \\
		\cite{pehle2022brainscales2_nopreprint_nourl,billaudelle2022accurate}					& \multicolumn{2}{l}{\cite{mueller2022scalable,spilger2020hxtorch,spilger2022hxtorchsnn}} \\
		\\
		ROLLS																					& physical                & bio.\ SNN \\
		\cite{qiao2015reconfigurable_nourl}														& \multicolumn{2}{l}{\cite{stefanini2014pyncs}} \\
		\\
		DynapSE																					& physical                & SNN \\
		\cite{moradi2018dynaps}																	& \multicolumn{2}{l}{\cite{dynapse2021nice}} \\
		\\
		Neurogrid																				& physical                & SNN \\
		\cite{benjamin2014neurogrid}															& \multicolumn{2}{l}{\cite{voelker2017extending}} \\
		\bottomrule
	\end{tabular}
	\caption{%
		Overview of neuromorphic chip architectures:
		At one end of the spectrum, programmable standard processors offer modeling flexibility that can come close to software simulations while maintaining higher efficiency.
		On the other end of the spectrum, analog circuits typically offer high energy efficiency or speed, but this is typically based on a trade-off in modeling flexibility or precision.
		Both digital and analog implementations offer many different optimization options.
		Some systems focus on biological \gls{snn} operation, while others support machine-learning-inspired training or classical ANN operation.
	}%
	\label{tab:neuromorphic_architectures}
\end{table*}

\newpage
\section{Spike time decoder}

\begin{listing}[H]
	\begin{minted}{python}
class ToSpikeTimes(torch.autograd.Function):

  def forward(
    ctx,
    spike_input: torch.Tensor,
    spike_count: torch.Tensor,
    dt: float) -> torch.Tensor:
  batch_size, time_size, out_size = spike_input.shape
  indexed_spike_input =
    spike_input * torch.arange(1, time_size + 1)[None, :, None] - 1.0
  indexed_spike_input[indexed_spike_input == -1.0] = float("inf")
  if spike_count < time_size:
    spike_indices = torch.sort(
		indexed_spike_input, dim=1).values[:, :spike_count]
  else:
    spike_indices = torch.sort(
      indexed_spike_input, dim=1).values[:, :time_size]
  ctx.save_for_backward(spike_indices, spike_input)
  ctx.shape = spike_input.shape
  return spike_indices * dt

  def backward(
      ctx,
      grad_output: torch.Tensor) -> Tuple[torch.Tensor, None, None]:
    (spike_indices, spike_input) = ctx.saved_tensors
    batch_size, spike_count, out_size = spike_indices.shape
    noninf_spike_indices = spike_indices.flatten() != float("inf")

    grad_input = torch.zeros_like(spike_input, dtype=torch.float)

    grad_input_indices = (
      torch.arange(batch_size).repeat_interleave(out_size).repeat(
        spike_count)[noninf_spike_indices],
      spike_indices.flatten()[noninf_spike_indices].type(torch.long),
      torch.arange(out_size).repeat(batch_size).repeat(
        spike_count)[noninf_spike_indices],
    )

    grad_input[grad_input_indices] = \
      - 1.0 * grad_output.flatten()[noninf_spike_indices]

    return grad_input, None, None
	\end{minted}
	\caption{%
		Custom \texttt{torch.autograd.Function} to convert boolean tensors holding spikes into spike times in a backpropagation-compatible fashion.
		The argument \texttt{spike\_count} adjusts how many of each neurons spikes are retrieved.
		If a neuron does spike fewer times than specified by \texttt{spike\_count}, the spike times are set to floating-point infinity. %
		The backpropagation happens by injecting the gradient with respect to a spike time at the corresponding position, at which the spike occurred in the boolean input tensor, with a negative sign.
		All other entries remain to be zero.
	}%
	\label{lst:to_spike_times}
\end{listing}

\newpage
\section{Training Parameters}
\begin{table}[htbp]
	\centering
	\begin{tabular}{l c c}
		\toprule
		\textbf{Parameter} & \quad \textbf{EventProp} & \quad \textbf{SuperSpike} \\
		\midrule
		$\tau_{\text{mem}}$ & \SI{6}{\micro\second} & \SI{6}{\micro\second} \\
		$\tau_{\text{syn}}$ & \SI{6}{\micro\second} & \SI{6}{\micro\second}\\
		 & \\
		size input & 5  & 5 \\
		size hidden layer & 120 & 120\\
		size output layer & 3 & 3\\
		weight init [mean, stdev] & & \\
		\hspace{4pt} hidden & [0.2, 0.2] & [0.001, 0.15] \\
		\hspace{4pt} output & [0.01, 0.1] & [0.0, 0.1] \\
		 & \\
		$dt$ & \SI{0.5}{\micro\second} & \SI{0.5}{\micro\second}\\
		$t_{\text{bias}}$ & \SI{2}{\micro\second} & \SI{2}{\micro\second} \\
		$t_{\text{early}}$ & \SI{2}{\micro\second} & \SI{2}{\micro\second} \\
		$t_{\text{late}}$ & \SI{26}{\micro\second} & \SI{26}{\micro\second} \\
		$t_{\text{sim}}$ & \SI{38}{\micro\second} & \SI{38}{\micro\second} \\
		 & \\
		 training epochs & 300 & 300 \\
		 batch size & 50 & 100 \\
		 optimizer & Adam & Adam \\
		 Adam parameter $\beta$ & (0.9, 0.999) & (0.9, 0.999) \\
		 Adam parameter $\epsilon$ & $10^{-8}$ & $10^{-8}$ \\
		 learning rate & 0.0005 & 0.001 \\
		 lr-scheduler & StepLR & StepLR \\
		 lr-schedule step size & 50 & 50 \\
		 lr-scheduler $\gamma$ & 0.5 & 0.5 \\
		 readout reg. & 0.0004 & 0.0004 \\
		\bottomrule
	\end{tabular}
	\caption{Parameters of neuron dynamics, network and training used to produce the hardware-in-the-loop results on the Yin-Yang dataset using EventProp and SuperSpike surrogate gradients as shown in \cref{fig:training_results_yinyang} and \cref{tab:result_comparison}.}%
	\label{tab:training_parameters}
\end{table}

\begin{table}[htbp]
	\centering
	\begin{tabular}{l c c}
		\toprule
		\textbf{Parameter} & \quad \textbf{EventProp} & \quad \textbf{SuperSpike} \\
		\midrule
		size input & 5  & 5 \\
		size hidden layer & 120 & 120\\
		size output layer & 3 & 3\\
		weight init [mean, stdev] & & \\
		\hspace{4pt} hidden & [1.0, 0.4] & [1.0, 0.4] \\
		\hspace{4pt} output & [0.01, 0.1] & [0.01, 0.1] \\
		 & \\
		$dt$ $\left[ \tau_{\text{s}} \right]$ & \SI{0.01}{} & \SI{0.01}{}\\
		$t_{\text{bias}}$ $\left[ \tau_{\text{s}} \right]$ & \SI{0}{} & \SI{0}{} \\
		$t_{\text{early}}$ $\left[ \tau_{\text{s}} \right]$ & \SI{0}{} & \SI{0}{} \\
		$t_{\text{late}}$ $\left[ \tau_{\text{s}} \right]$ & \SI{4}{} & \SI{4}{} \\
		$t_{\text{sim}}$ $\left[ \tau_{\text{s}} \right]$ & \SI{6}{} & \SI{6}{} \\
		 & \\
		 training epochs & 200 & 200 \\
		 batch size & 25 & 50 \\
		 optimizer & Adam & Adam \\
		 Adam parameter $\beta$ & (0.9, 0.999) & (0.9, 0.999) \\
		 Adam parameter $\epsilon$ & $10^{-8}$ & $10^{-8}$ \\
		 learning rate & 0.0005 & 0.0005 \\
		 lr-scheduler & StepLR & StepLR \\
		 lr-schedule step size & 50 & 50 \\
		 lr-scheduler $\gamma$ & 0.5 & 0.5 \\
		 readout reg. & 0.0 & 0.0 \\
		\bottomrule
	\end{tabular}
	\caption{Parameters of neuron dynamics, network and training used to produce the software-only results on the Yin-Yang dataset using EventProp and SuperSpike surrogate gradients as listed \cref{tab:result_comparison}. We chose $\tau_{\text{s}} = \tau_{\text{m}}$ and times are given in units of $\tau_{\text{s}}$.}%
	\label{tab:training_parameters_mock}
\end{table}

\newpage
\section{EventProp Dataflow}

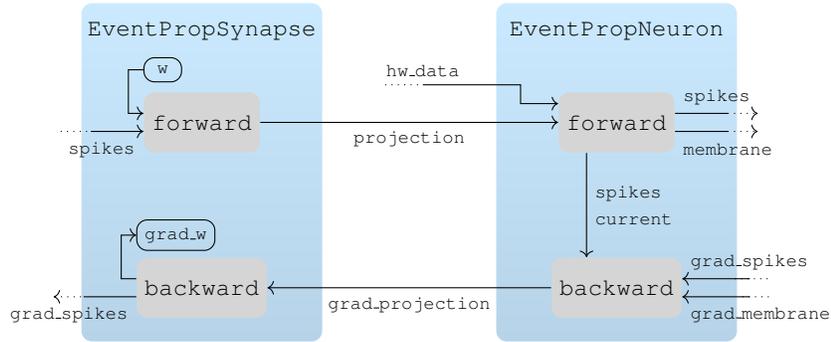
\begin{figure*}[htbp]
	\centering
	\definecolor{dataflow_blue}{HTML}{d5ecfd}  %
\definecolor{dataflow_grey}{HTML}{d5d5d5}  %
\definecolor{upper_blue}{HTML}{cbeaff} %
\definecolor{lower_blue}{HTML}{b6d3e8} %

\begin{tikzpicture}
\begin{footnotesize}

\node[top color=upper_blue, bottom color=lower_blue, minimum height=4.5cm, minimum width=3.2cm, rounded corners] (syn_mod) {};
\node[above=-0.6cm of syn_mod] (synapse_label) {\texttt{EventPropSynapse}};
\node[above=-2cm of syn_mod, fill=dataflow_grey, minimum height=0.8cm, minimum width=1.4cm, rounded corners] (syn_fwd) {\texttt{forward}};
\node[anchor=west, minimum height=0.3cm, minimum width=0.5cm, draw=black, rounded corners] (w) at ($(syn_fwd.west)+(0, 0.7cm)$)  {\scriptsize \texttt{w}};
\draw[->] (w.west) -- ($(w.west)+(-0.2cm,0)$) -- ($(syn_fwd.west)+(-0.2cm, 0.12cm)$) -- ($(syn_fwd.west)+(0, 0.12cm)$);
\node[above=-4.2cm of syn_mod, fill=dataflow_grey, minimum height=0.8cm, minimum width=1.4cm, rounded corners] (syn_bwd) {\texttt{backward}};
\node[anchor=west, minimum height=0.3cm, minimum width=0.6cm, draw=black, rounded corners] (grad_w) at ($(syn_bwd.west)+(0, 0.7cm)$)  {\scriptsize \texttt{grad\_w}};
\draw[<-] (grad_w.west) -- ($(grad_w.west)+(-0.2cm,0)$) -- ($(syn_bwd.west)+(-0.2cm, 0.12cm)$) -- ($(syn_bwd.west)+(0, 0.12cm)$);

\node[right=2.3cm of syn_mod, top color=upper_blue, bottom color=lower_blue, , minimum height=4.5cm, minimum width=3.2cm, rounded corners] (nrn_mod) {};
\node[above=-0.6cm of nrn_mod] (neuron_label) {\texttt{EventPropNeuron}};
\node[above=-2cm of nrn_mod, fill=dataflow_grey, minimum height=0.8cm, minimum width=1.4cm, rounded corners] (nrn_fwd) {\texttt{forward}};
\node[above=-4.2cm of nrn_mod, fill=dataflow_grey, minimum height=0.8cm, minimum width=1.4cm, rounded corners] (nrn_bwd) {\texttt{backward}};

\draw[->] ($(syn_fwd.west)+(-0.7cm, -0.12cm)$) -- ($(syn_fwd.west)+(0, -0.12cm)$);
\draw[dotted] ($(syn_fwd.west)+(-1.1cm, -0.12cm)$) -- ($(syn_fwd.west)+(-0.7cm, -0.12cm)$);
\node[anchor=east] at ($(syn_fwd.west)+(0, -0.35cm)$) {\scriptsize \texttt{spikes}};

\draw[->] (syn_fwd.east) -- (nrn_fwd.west) node[below, midway] {\scriptsize \texttt{projection}};

\draw[->] ($(nrn_fwd.south)+(-0.4cm, 0)$) -- ($(nrn_bwd.north)+(-0.4cm, 0)$) node[right, midway, text width=1.5cm] (save_spikes) {\scriptsize \texttt{spikes} \scriptsize \texttt{current}};

\draw ($(nrn_fwd.east)+(0, 0.12cm)$) -- ($(nrn_fwd.east)+(0.7, 0.12cm)$);
\draw[->, dotted] ($(nrn_fwd.east)+(0.7, 0.12cm)$) -- ($(nrn_fwd.east)+(1.1, 0.12cm)$);
\node[anchor=west] at ($(nrn_fwd.east)+(0, 0.35cm)$) {\scriptsize \texttt{spikes}};
\draw ($(nrn_fwd.east)+(0, -0.12cm)$) -- ($(nrn_fwd.east)+(0.7, -0.12cm)$);
\draw[->, dotted] ($(nrn_fwd.east)+(0.7, -0.12cm)$) -- ($(nrn_fwd.east)+(1.1, -0.12cm)$);
\node[anchor=west] at ($(nrn_fwd.east)+(0, -0.35cm)$) {\scriptsize \texttt{membrane}};

\draw[<-] ($(nrn_bwd.east)+(0, 0.12cm)$) -- ($(nrn_bwd.east)+(0.9, 0.12cm)$);
\draw[dotted] ($(nrn_bwd.east)+(0.7, 0.12cm)$) -- ($(nrn_bwd.east)+(1.2, 0.12cm)$);
\node[anchor=west] at ($(nrn_bwd.east)+(0, 0.35cm)$) {\scriptsize \texttt{grad\_spikes}};
\draw[<-] ($(nrn_bwd.east)+(0, -0.12cm)$) -- ($(nrn_bwd.east)+(0.9, -0.12cm)$);
\draw[dotted] ($(nrn_bwd.east)+(0.8, -0.12cm)$) -- ($(nrn_bwd.east)+(1.2, -0.12cm)$);
\node[anchor=west] at ($(nrn_bwd.east)+(0, -0.35cm)$) {\scriptsize \texttt{grad\_membrane}};

\draw[<-] (syn_bwd.east) -- (nrn_bwd.west) node[below, midway] {\scriptsize \texttt{grad\_projection}};

\draw ($(syn_bwd.west)+(-0.7cm, -0.12cm)$) -- ($(syn_bwd.west)+(0, -0.12cm)$);
\draw[<-, dotted] ($(syn_bwd.west)+(-1.1cm, -0.12cm)$) -- ($(syn_bwd.west)+(-0.7cm, -0.12cm)$);
\node[anchor=east] at ($(syn_bwd.west)+(0, -0.35cm)$) {\scriptsize \texttt{grad\_spikes}};

\draw[dotted] ($(nrn_fwd.west)+(-2.3cm, 0.5cm)$) -- ($(nrn_fwd.west)+(-1.8cm, 0.5cm)$);
\draw[->] ($(nrn_fwd.west)+(-1.8cm, 0.5cm)$) node[above] {\scriptsize \texttt{hw\_data}} -- ($(nrn_fwd.west)+(-0.5cm, 0.5cm)$) -- ($(nrn_fwd.west)+(-0.5cm, 0.25cm)$) -- ($(nrn_fwd.west)+(0, 0.25cm)$);

\end{footnotesize}
\end{tikzpicture}
	\caption{%
		Dataflow occurring during hardware-in-the-loop training. \texttt{EventPropSynapse} returns the product of the input spikes and the weight $w$ stacked on top of an empty tensor with same shape.
		\texttt{EventPropNeuron} either returns the hardware observations or, if those are not present, computes the forward trajectories in simulation.
		In the backward pass, the adjoint dynamics are computed using the stored spikes from the \texttt{forward}-call and a stacked tensor consisting of $\tau_{\text{s}} \lambda_{\text{I}}$ and $(\lambda_{\text{I}} - \lambda_{\text{V}})$ is returned.
		Note that returning these stacked tensors is only possible due to the output of \texttt{EventPropSynapse} already being a stacked tensor with same shape.
		In backward direction, \texttt{EventPropSynapse} backpropagates $\tau_{\text{s}} \lambda_{\text{I}}^{\top} z_{\text{pre}}$, the gradient with respect to the weight according to the EventProp algorithm~\cite{wunderlich2021event}, and $(\lambda_{\text{I}} - \lambda_{\text{V}}) w$, the gradient with respect to the pre-synaptic spikes.
	}%
	\label{fig:pytorch_modules}
\end{figure*}

\end{document}